\documentclass[12pt,preprint]{aastex}

\begin{document}

\title {DISPARITY BETWEEN H$\alpha$ AND H$\beta$ IN SN~2008in:
INHOMOGENEOUS EXTERNAL LAYERS OF TYPE IIP SUPERNOVAE?
} 

\bigskip

\author{N. N.~Chugai \altaffilmark{1}, 
V. P. Utrobin \altaffilmark{2}}
\altaffiltext{1}{Institute of astronomy RAS, Moscow 119017, Pyatnitskaya 48;
nchugai@inasan.ru}
\altaffiltext{2}{Institute of Theoretical and Experimental Physics, 
Moscow 117218, B. Cheremushkinskaya st. 25; utrobin@itep.ru}

\begin{abstract}

We study disparity between  H$\alpha$ and H$\beta$ in early spectra 
of the type IIP supernova SN~2008in. The point is that these lines cannot 
be described simultaneously in a spherically-symmetric model with the smooth 
density distribution. It is shown that an assumption of a clumpy structure 
of external layers of the envelope resolves the  
problem. We obtain estimates of the velocity at the inner border of 
the inhomogeneous zone ($\approx6100$ km s$^{-1}$), the filing factor 
of inhomogeneities ($\leq0.5$), and the mass of the inhomogeneous layers 
($\sim0.03~M_{\odot}$). The amplitude of flux fluctuations in the early 
spectrum of H$\alpha$ ($\Delta F/F\sim10^{-2}$) imposes a constraint on 
the size of inhomogeneities ($\leq 200$ km s$^{-1}$). A detection of 
fluctuations in the early H$\alpha$ of type IIP supernovae might become 
an observational test of the inhomogeneous structure of 
their envelopes. We propose also the indirect test of the  
clumpy structure of external layers: the study of properties 
of the initial radiation outburst due to the shock breakout.
The inhomogeneous structure of external layers of type IIP 
supernovae could be an outcome of density perturbations and density 
inversion in outer convective layers of presupernova red supergiant.

\end{abstract}

Keywods: {\it supernovae and supernova remnants; stars --- structure and 
evolution}

\bigskip

\section{Introduction}

Type IIP supernovae (SN~IIP) represent majority of core-collapse 
events in the mass range of $9-25~M_{\odot}$ (Heger et al. 2003).
Despite the explosion mechanism is not well understood, the major 
observational characteristics of SN~IIP indicate that we deal with 
the red supergiant (RSG) explosion (Grassberg et al. 1971; Falk \& Arnett 1977; 
Eastman et al. 1994; Baklanov et al. 2005; Utrobin 2007).
Yet detailed modelling of eight well observed SN~IIP uncovers a 
serious problem that progenitor masses of all of them 
turn out to be significantly larger $M_{psn}\geq15~M_{\odot}$ compared to 
the lower limit ($\approx9~M_{\odot}$) of massive stars 
(Utrobin \& Chugai 2013, henceforth UC13). Although this 
sample is small and selection effects might skew the distribution, 
this fact poses a difficult question, whether we understand adequately 
the SN~IIP light curve and spectrum formation.

Recently, modelling the type IIP supernova SN~2008in revealed 
an unexpected problem, namely, that H$\alpha$ and H$\beta$ line profiles 
cannot be described 
simultaneously in the spherically-symmetric model (UC13).
Specifically, H$\beta$ absorption turns out to be substantially deeper 
compared to the prediction based on the model adjusted to H$\alpha$.
The problem seems to be rather serious keeping in mind simplicity of 
the hydrogen spectrum and the fact that the Balmer lines have the common 
lower level. In that paper we suggested that the inconsistency between 
H$\alpha$ and H$\beta$ could be an outcome of an inhomogeneous 
structure of supernova envelope in outer layers.

Here we explore a possibility for the solution of the H$\alpha$/H$\beta$ 
problem in SN~2008in by assuming the inhomogeneous (clumpy) matter 
distribution in outer layers of supernova.
We confine ourselves to the question, whether one could find 
an acceptable description of H$\alpha$ and H$\beta$ profiles
assuming a clumpy atmosphere, and, if so, what should be 
the optimal parameters of the inhomogeneous structure.
We start with a description of H$\alpha$/H$\beta$ problem using 
the model with a smooth density distribution (section 2). We then
present the inhomogeneous model and results of the modelling for 
H$\alpha$ and H$\beta$ line profiles (section 3). In section 4 
we consider the issue of flux fluctuations in the H$\alpha$ line 
related to the inhomogeneous structure.

Our study is based on spectra of SN~2008in published by Roy et al. (2011).

\section{H$\alpha$/H$\beta$ problem}

We consider a freely expanding spherical envelope with the Lagrangian
velocity and radius obeyed the relation $v=r/t$, where $t$ is the time 
since the explosion. A photosphere is presumably sharp and resides at 
the level $v_p$; above that level the atmosphere is transparent in 
the continuum. For the moderate H$\alpha$ optical depth  
($\tau_{23}<10^5$) the size of the scattering region of a photon is 
of the order of the sound scale length $l_s\sim u t$, 
($u$ is the hydrogen thermal velocity).
Given $l_s\ll r\sim v_pt$, the Sobolev approximation for the local photon 
scattering is valid with the high precision. The line profile formed in 
the supernova envelope is determined by the behavior of the 
local optical depth $\tau(v)$ and source function $S(v)$. 
Since we are interested in the relative flux, 
in case of the uniform brightness $I^c$ of the photosphere 
the source function can be written in the dimensionless 
form $S=W+S_e$, where $W$ is the dilution 
factor. The first term is for the scattering of photospheric 
radiation and the second term is for the intrinsic emission.
The power law is adopted for the optical depth $\tau(v)=\tau_p(v/v_p)^p$ 
and for the source function $S_e(v)=S_{e,p}(v/v_p)^q$, where 
$\tau_p$ and $S_{e,p}$ are values at the photosphere.

In Fig. 1 we show two versions of models for 
SN~2008in on day 11. The first one is aimed 
at reproducing H$\alpha$, whereas the second at reproducing 
H$\beta$. The photospheric velocity is 6100 km s$^{-1}$ in both cases.
The plots show that the model being good for the H$\alpha$ is bad 
for H$\beta$. Two possibilities for H$\beta$ are shown: without and with 
the net emission. In both cases the fit is bad.
On the other hand, the model aimed at the description of H$\beta$ 
does not reproduce H$\alpha$. 
This modelling demonstrates the core of the H$\alpha$/H$\beta$ problem 
in the early spectrum of SN~2008in: the lines cannot be reproduced 
simultaneously in the model of the smooth spherically-symmetric envelope.
To put it staightforwardly, the H$\beta$ absorption is stronger 
than that in the model based on H$\alpha$. The opposite is true as well: the 
H$\alpha$ absorption is weaker than that in the model based on the H$\beta$.
It should be emphasized that this controversy cannot be resolved 
by taking metal lines into account because at this stage metal lines 
are very weak and cannot markedly affect hydrogen line profiles.

We explored the effects of a multiple Thomson scattering off 
thermal electrons taking into account the H$\alpha$ and H$\beta$ 
photons created at large optical depth $\tau_{\rm T} \leq 3$. 
This mechanism is known to produce broad wings in Balmer lines of 
some SN~IIn (Chugai 2001). Our modelling shows however that for reasonable 
temperature ($\sim10^4$ K) the effects of wings are negligible and 
cannot resolve the H$\alpha$/H$\beta$ problem anyway.

\section{Inhomogeneous envelope}

Arguments for the consideration of the inhomogeneous model stem
from the fact that in the clumpy medium with the 
large cloud to intercloud contrast the line absorption depth is determined 
by both atomic constants and the cloud filling factor. The situation 
is conceivable when the H$\alpha$ and H$\beta$ absorptions in clouds 
are saturated. In this case, if the intercloud gas is absent,  
the H$\alpha$/H$\beta$ absorption ratio could become close to 
unity, i.e., significantly lower than the homogeneous model predicts. 

\subsection{Overview of inhomogeneous model}

We suggest that the supernova atmosphere consists of an ensemble 
of dense clouds embedded in a rarefied intercloud medium. The 
inhomegeneous  structure is characterized by the volume filling factor 
$f=NV_c/L^3$, where $N$ is the average cloud number in a volume 
$L^3$, $V_c$ is the average volume of a cloud. Generally, $f$ is 
a function of the Lagrangian radius. If a cloud size $l_c$ 
much larger than the size of a resonance region (or sound radius) 
$l_c\gg l_s$, the average total area of sections of clouds by 
a random plane (resonance plane in our case) with the area $\Delta A$
is equal $f\Delta A$ according to the well-known theorem of geometrical 
probability (Kendal \& Moran 1963). With this result,
the intensity of the escaping photospheric radiation 
(neglecting emission component) at the radial velocity $v_z$, 
i.e., at the frequency $\nu=\nu_0(1-v_z/c)$ is
\begin{equation}
I_{\nu} = I_{\nu}^c[f\exp\,(-\tau_c)+(1-f)\exp\,(-\tau_{ic})]\,,
\label{eq-ffac}
\end{equation}
where $\tau_c$ is the line optical depth of the cloud 
and $\tau_{ic}$ is the optical depth in the intercloud 
medium. The expression in square brackets has a meaning 
of the attenuation factor; it depends on the frequency and the radius.
In the limit $l_c\gg l_s$ this factor does not depend on the cloud size.
In case of an emission component the effect of clumpiness can be calculated 
similar to the absorption component.

As we will see below, to reproduce Balmer lines, the 
inhomogeneity should be invoked only in the outermost layers. We therefore 
consider the
smooth matter distribution ($f=1$) in the inner envelope zone $v < v_f$ 
and the clumpy distribution in the outer zone ($v \geq v_f$) assuming a power 
law for the radial variation
\begin{equation}
f=f_0(v/v_f)^{k}\,.
\label{eq-filf}
\end{equation}

In case of small clouds ($l_c\sim l_s$) the relation (\ref{eq-ffac}) 
is not generally applicable. Using this expression would result in the 
error which can be estimated by the Monte Carlo (MC) technique.
To this end we adopt a random distribution of spherical clouds with a 
radius $a$ in the envelope with the kinematics $v=r/t$. 
For $a=5l_s$ we find that the attenuation factor calculated by MC
coincides with the value found from the equation (\ref{eq-ffac}).
In case of $a=3l_s$ the 
analytic formula gives the attenuation factor 8\% larger than MC simulation;
for $a=l_s$ the difference is 23\%. The analytic formula therefore 
is applicable in a broad range of cloud sizes $l_c>3l_s$.

\subsection{Modelling results}

We start with the constant filling factor $f=f_0$
(i.e., $k=0$) for $v>v_f$. In this case to sensibly describe  
Balmer lines in SN~2008in on day 11, the whole atmosphere 
 should be inhomogeneous, $v_f\approx v_p=6100$ km s$^{-1}$. The 
optimal value of the filling factor should be $f=0.4$ to fit the H$\beta$
(Fig. 2b). Yet the model $f=$const fails to adequately describe 
simultaneously both the H$\alpha$ and H$\beta$ lines (Fig 2). The shown 
model is transparent in the intercloud medium. Introducing a finite 
absorption in the 
intercloud gas worsen the agreement. We conclude therefore that the 
model with constant filling factor should be abandoned.

Assuming a variable filling factor permits us to easily come to
the optimal model describing both Balmer lines 
(Figs. 2c,d, model D11 in Table).
Table contains the velocity at the photosphere derived from 
H$\alpha$ and H$\beta$ [$v_p(\rm{H}\alpha)$ and $v_p(\rm{H}\beta)$],
the velocity at the inner boundary of the inhomogeneous zone ($v_f$), 
the optical depth of clouds ($\tau_c$) and intercloud gas ($\tau_{ic}$) 
in the H$\alpha$ at $v=v_f$, power law index $p$ in the $\tau\propto v^p$ 
relation, power law index $q$ in the $S\propto v^q$ relation, and power 
law index $k$ in the equation (\ref{eq-filf}).
An interesting point of our results is the relatively weak radial
dependence of the population of the hydrogen second level 
$n_2\propto\tau_{23}\propto v^{-5.5}$ compared to total 
density $\rho\propto v^{-7.6}$ in the hydrodynamic model of SN~2008in 
(UC13). A similar remark refers to the flat behavior 
of the source function $S_e=$const. Both facts indicate the 
enhanced excitation 
of hydrogen compared to the case when the excitation by the
photospheric radiation dominates.

The inhomogeneous zone in our model extends as deep as the photosphere 
$v_f\approx v_p=6100$ km s$^{-1}$. This poses a question, whether the 
inhomogeneous structure extends to deeper layers. As to the situation on 
day 11, the density distribution of the SN~2008in
hydrodynamic model (UC13) suggests that the mass of the 
inhomogeneous atmosphere ($v>6100$ km s$^{-1}$) turns out to be 
$\approx0.03~M_{\odot}$. This is only $\sim2\times10^{-3}$
of the total mass of ejecta ($13.6~M_{\odot}$). 
The mass of the inhomogeneous zone is a crucial observational constraint 
on the mechanism of the inhomogeneities formation 
in the outer layers of SN~IIP.

\subsection{Evolution effects}

The possible clumpy structure in the inner layers 
($v<6100$ km s$^{-1}$) can be studied using the spectrum of SN~2008in 
on day 18 (Roy et al. 2011) when the photosphere is deeper than 
on day 11. We thus can check whether the inhomogeneous 
structure at $v<6100$ km s$^{-1}$ is needed in order to describe 
H$\alpha$ and H$\beta$ at this stage. First, however, we answer the question
whether the H$\alpha$/H$\beta$ problem retains on day 18 in the 
homogeneous model.

Unlike previous epoch, the spectrum on day 18 shows rather strong metal 
lines which could affect hydrogen line profiles. We therefore should calculate 
the spectrum taking into account metal lines. 
The list of $\approx1.9\times10^4$ lines in the spectral 
range 3500-10000~\AA\ for the condition of supergiant 
with the temperature of 7000~K has been retrieved from the VALD 
database 
(Kupka et al. 1999). From this list we select lines in the range 
of H$\alpha$ and H$\beta$ with the optical depth $\tau>0.3$ at the 
photosphere for the excitation temperature of 7000~K. We assume that 
in metal lines the radiation 
is scattered with a finite absorption probability ($\sim 10^{-2}$) 
taken to be the same for all metal lines. In the H$\alpha$ and H$\beta$ 
lines we take into account both scattering and emission. Level populations 
are not in equilibrium which poses some problem. Fortunately, in 
the H$\alpha$ band  
Si\,II 6347, 6371 \AA\ doublet dominates,  while in the H$\beta$ 
band Fe\,II multiplet 42 
prevails which facilitates a search for the optimal model. Ion abundances
for adopted excitation temperature (7000~K) are adjusted using different 
correction factors for iron peak elements, Si, and hydrogen. 

The modelling shows (Fig. 3)  that to reproduce the H$\beta$ line,
the hydrogen second level population should be larger by a factor of 
three than that 
recovered from the H$\alpha$. Remarkably, the similar factor value is needed 
to reproduce the H$\alpha$ and H$\beta$ lines on day 11 using the smooth model 
(UC13). The modelling demonstrates that the problem of the
H$\alpha$/H$\beta$ disparity retains on day 18 in the model of 
smooth density distribution. The question left to answer is 
whether the clumpy model for day 11 is applicable to day 18. 

First, however, we comment an interesting fact which was missing 
formerly in available analysis of SN~IIP spectra. The point is that on day 18,
unlike the earlier phase, the velocity at the photosphere measured from
the H$\alpha$ and H$\beta$ turns out to be different: 
 4200 km s$^{-1}$ and 4700 km s$^{-1}$ respectively.
At first glance the value derived from the H$\beta$ with a low net emission 
is more confident than that 
derived from the H$\alpha$. Moreover, Fe\,II 4924, 5018, 5169 \AA\ lines 
(multiplet 42) confirm the value found from H$\beta$ (Fig. 3a). 
On the other hand, 
the velocity of 4200 km s$^{-1}$ obtained from H$\alpha$ is consistent 
with the 
Si\,II 6347, 6371 \AA\ doublet (Fig. 3b) which also rules out larger velocity 
($>4300$ km s$^{-1}$). This fact indicates that the photosphere indeed 
lies at different 
levels in the H$\alpha$ and H$\beta$ bands: it is deeper for the H$\alpha$ 
than for the H$\beta$. We believe that this difference is related 
to the fact that the contribution of metal lines in the quasi-continuum 
is larger in the  H$\beta$ band compared to the H$\alpha$ band. 
The velocity at the photosphere around day 20 in these bands could 
become a valuable constraint on the 
model of SN~IIP based on the multi-group radiation transfer.

We turn now to the modelling of the H$\alpha$ and H$\beta$ on day 18 taking 
into account inhomogeneities in the envelope. The same model 
of the inhomogeneous structure is used 
as on day 11. Parameters of the optimal model D18 are given in Table. 
The calculated profiles for this model fit the observations well enough. 
In this model the photospheric velocities
in the H$\alpha$ and H$\beta$ lines are 4100 km s$^{-1}$ and 5300 km s$^{-1}$
respectively which qualitatively agree with the velocities found in 
the smooth model. Remarkably, parameters of the inhomogeneous structure 
seem to be the same on days 11 and 18 with the exception that the
intercloud optical depth on day 18 is a bit lower, 1.5 instead of 2, 
which is in line with the excitation decrease demonstrated by the spectra.
The fact that parameters of the inhomogeneous structure 
did not change noticeably means that inhomogeneity scales
evolve homologously $l\propto r$ which in turn indicates the
early formation of the inhomogeneous structure perhaps
at the shock breakout stage. 

\section{Inhomogeneous structure and flux fluctuations in H$\alpha$}

The H$\alpha$ emission in the inhomogeneous model originates from
dense clumps. One should expect therefore the presence of flux 
fluctuations in the emission component. The detection of these fluctuations 
could become a test for the inhomogeneous structure. 
A rough estimate 
of the fluctuation amplitude can be obtained assuming the Poisson distribution 
of clouds. In this case the amplitude of the flux fluctuations in the range of 
radial velocities $\Delta v$ is equal $\Delta F/F\sim N^{-1/2}$, where 
$N$ is the number of clouds that contribute to the line radiation flux $F$ in 
this velocity range 
\begin{equation}
N\approx (3/4)f(v_pt/a)^2(\Delta vt/a)\,.
\label{eq-fluc}
\end{equation}
Here $f$ is the volume filling factor of clouds, $a$ is the 
average cloud radius, 
$v_p$ is the velocity at the photosphere. In the observed 
spectrum on day 11 the 
amplitude of flux fluctuations in the H$\alpha$ profile is approximately 
$\Delta F/F\sim 1$\%. The number of clouds in the layer with the width 
corresponding to the spectral resolution  
$\Delta v\approx300$ km s$^{-1}$ should be 
thus $N\approx10^4$. With this value the equation (\ref{eq-fluc}) 
results in the cloud radius $a\sim60$ km s$^{-1}$ assuming 
$f\sim0.3$ and $v_p=6100$ km s$^{-1}$. The characteristic size of
inhomogeneities is thus $\sim2a\sim10^2$ km s$^{-1}$, if the fluctuations 
are actually related to the inhomogeneous structure. 

The Poisson distribution however overestimates the fluctuation amplitude 
because it admits the cloud overlapping. This hardly occurs in reality 
since the inhomogeneities usually form in the process of fragmentation 
which basically suggests divergent flows. In order to avoid overlapping,
we use the hard sphere model for the clouds;
a similar model was assumed for the analysis of fluctuations in 
the [O\,I] 6300 \AA\ doublet in SN 1987A (Chugai 1994). The cloud 
distribution is realised in several steps. First we place a sphere 
of a radius $a$ 
randomly in a cubic cell of the size $b=a(4\pi/3f)^{1/3}$ with the 
filling factor $f$ being the function of velocity $f(v)\propto v^{-2}$.
Cubic cells are placed in a spherical layer of a radius $r$ and the 
thickness of $b$ in the following way. The cells are placed one by one
in a ring element 
$2\pi r^2b\sin\,\theta\,d\theta$ (where $\theta$ is the polar angle 
measured from $z$-axis which directed along the line of sight).
In each other ring the initial random shift was introduced. Each filled 
spherical layer is rotated on the random angle around $y$-axis which 
lies in the sky plane. The overall result is a random angular distribution of 
non-overlapping spherical clouds in the envelope. In the radial 
distribution the 
randomness presents only in the radial position of a sphere in each cell.
The lack of a full chaos along the radius does not affect 
the result because we are interested only in a
one-dimensional projection of the three-dimensional velocity space 
on the $z$-axis in a narrow interval around zero radial velocity 
($|v_z|<v_p$). In this case the flux fluctuations in the H$\alpha$ are 
not sensitive to the radial distribution of clouds.

With the obtained random distribution of spherical clouds in the 
supernova envelope we performed a set of computations for the velocity 
of 6100 km~s$^{-1}$  at the photosphere adopting the 
behavior of the source function and optical depth in clouds according to 
those found on day 11. The derived line profile is convolved with the 
Gaussian adopting the full width at half maximum of 300 km s$^{-1}$ to allow
for the spectral resolution. We show in Fig. 5a the calculated 
residual spectrum 
$\Delta F/F=F/F_{sm}-1$ (where $F_{sm}$ is heavily smoothed version of 
$F$) for two cases $a=100$ km s$^{-1}$ and $a=250$ km~s$^{-1}$.
The comparison of these residuals with that for the observed spectrum of
SN~2008in on day 11 (Fig. 5b) permits us to constrain 
the scale of inhomogeneities. It is obvious that the inhomogeneities 
substantially larger than $2a\sim 200$ km s$^{-1}$ (e.g., 500 km s$^{-1}$)
are rulled out because they predict too large fluctuations.
The data do not rule out the cloud scales of $\sim 200$ km s$^{-1}$.
Note that the observed fluctuations include not only fluctuations related to 
inhomogeneities but also other noise sources (e.g., photon noise, 
background noise, and device noise). The expected fluctuations related 
to the envelope inhomogeneous structure therefore are lower than 1\%.
This suggests that an inhomogeneity scale should not exceed 200 km s$^{-1}$.

The model of spherical clouds is an oversimplification of 
the reality. It may well be that a significant fraction of inhomogeneities 
forms due to the corrugation of the shock front and subsequent shock
focusing on density perturbations. In that case one expects the 
creation of inhomogeneities with a filamentary geometry. Such 
inhomogeneities should be characterized by two essentially different scales.
The inhomogeneities with the filamentary structure could be 
generated also by Richtmyer-Meshkov, Rayleigh-Taylor, and 
Kelvin-Helmholtz instabilities which can develop during the shock 
propagation in the medium 
with the density inversion and density perturbations in the 
RSG atmosphere. The issue of the flux fluctuations in the H$\alpha$ 
profile thus may turn out to be a more complicated than in our simple 
model. Particularly, the filamentary structure could result in 
a second scale related to the filament length in the fluctuation spectrum.

\section{Discussion}

The paper has been aimed at the study of the disparity between 
the H$\alpha$ and H$\beta$ in early spectra of SN~2008in. The 
point is that in a standard model with a smooth density distribution 
the density required to reproduce the H$\beta$ line is by a factor of 
three larger
than that required for H$\alpha$. We find that this problem arises 
not only on day 11 but on day 18 as well. Assuming inhomogeneous 
(clumpy) density distribution in the outer layer of the supernova envelope,
we are able to describe H$\alpha$ and H$\beta$ profiles using the same model 
on days 11 and 18. This suggests homologous evolution of clumpy
structure, which is expected, if inhomogeneities form at the early expansion
phase.

The hydrodynamic model of SN~2008in permits us to estimate the 
mass of the external clumpy zone ($v\geq6100$ km s$^{-1}$) which turns 
out to be $\approx0.03~M_{\odot}$, i.e., only 0.2\% of the ejecta mass. 
Confronting the flux fluctuations in the model H$\alpha$ profile against the 
observed fluctuations on day 11 ($\leq 1$\%) suggests the upper limit 
of the inhomogeneities size to be  $\sim 200$ km s$^{-1}$. This value 
should be closely related to a process responsible for the formation 
of the inhomogeneous 
structure. In order to detect the fluctuations related to the inhomogeneities,
the spectra in the H$\alpha$ band should be obtained with the signal-to-noise 
ratio of $\approx200$ and the spectral resolution of $\approx 100$ km~s$^{-1}$.

We suggest that the formation of the inhomogeneous structure 
is related to specific conditions of the shock propagation in the 
RSG external layers with a mass of (several)$\times10^{-2}~M_{\odot}$. 
It is remarkable, in this respect, that RSGs are known to have 
a density inversion at the outer boundary of convective zone 
(Maeder 1992; Chiavassa et al. 2011). The depth of the 
density inversion in RSGs with masses of
$10... 20~M_{\odot}$ corresponds to $0.003... 0.1~M_{\odot}$ of external 
mass (Fadeev 2013, private communication). It is noteworthy that the mass of 
heterogeneous layers of SN~2008in ($\sim0.03~M_{\odot}$) falls into 
that range. The propagation of the radiation-dominated shock across 
the density inversion should be accompanied by the formation of a thin 
swept-up shell and its subsequent fragmentation. 

Another mechanism for the 
inhomogeneous structure could be related to density perturbations 
produced by a vigorous convection. 
In the RSG outer layers the velocity of convective flow is comparable with 
the sound speed $v\sim c_s$ (Maeder 1992). Since the density
perturbations are of the order of $\Delta \rho/\rho\sim (v/c_s)^2$ 
(Landau \& Lifshitz 1987), the expected perturbations can 
be as large as $\Delta \rho/\rho\sim 1$. The shock propagation 
in the medium with that strong density perturbations should result in the 
formation of inhomogeneities with the large density contrast. 
Laboratory experiments on the shock propagation in a turbulent medium
demonstrate a significant increase of amount of inhomogeneities with 
the increase of the Mach number (Hesselink \& Sturtevant 1988).
Disregarding the H$\alpha$/H$\beta$ problem,
it is little doubt that the inhomogeneous structure of the outer layers of 
SN~IIP should emerge anyway due to the shock wave propagation in 
the convective zone of the RSG. Moreover, inhomogeneities can be
generated additionally by the shock interaction with 
the RSG density inversion layer.

An interesting outcome of the inhomogeneous structure of the 
external layers could be a modification (compared to the
homogeneous case) of the initial radiation outburst of SN~IIP 
related to the shock
breakout. Indeed, the radiation diffusion time depends on the characteristics
of inhomogeneities in addition to absorption and scattering coefficients. 
This issue is of significant importance in view of the search for the initial 
ultraviolet radiation outburst from core-collapse SNe at the high redshift.

Another interesting consequence of the inhomogeneous structure of outer 
layers of SN~IIP concerns the circumstellar interaction.
The ejecta inhomogeneities could modify a standard picture of 
the direct and reverse shock. Instead of a spherical 
reverse shock a complicated system of the intercloud reverse 
shock and slow cloud  shock should emerge.
Moreover, the structure of the direct shock may be modified by 
protrusions created by dense ejecta clumps. All these features 
could affect the properties of radio, X-ray, and optical radiation 
of SN~IIP (and possibly SN~IIL) related to the circumstellar interaction.
 
To what extent the H$\alpha$/H$\beta$ problem is ubiquitous for SN~IIP? 
In case of SN~2006bp (type IIP), for which a good set of spectra has been 
obtained at the early epoch (Quimby at al. 2007), the synthetic spectra 
that describe the H$\beta$ at 8-16 days predict too deep 
H$\alpha$ absorption compared to observations (cf. Dessart et al. 2008).
The early spectra of SN~2006bp and SN~2008in 
look similar: the H$\beta$ absorption is strong whereas the 
H$\alpha$ absorption is absent. A similar picture is seen in early spectra 
of other normal SN~IIP, particularly, SN~1999em (Leonard et al. 2002) 
and SN~2012A (Tomasella et al. 2013). Remarkably, the modelling of 
the spectrum of SN~1993J (type IIb) on day 10 with the PHOENIX code 
(Baron et al. 1995) shows the same problem: the strong model H$\alpha$ 
absorption for the sensibly fitted H$\beta$ absorption (Fig. 5 in the 
referred paper). It should be noted that the SN~1993J pre-SN was a yellow 
supergiant (Aldering \& Humphreys 1994). On the other hand, 
for SN~1987A, in which case the pre-SN was a blue supergiant, the 
disparity between H$\alpha$ and H$\beta$ is absent (UC13). This 
fact supports our conjecture that the H$\alpha$/H$\beta$ problem 
arises only in the case of explosions of red (yellow) supergiants and not 
blue supergiants.

\section{Conclusions}

We demonstrate that the problem of the H$\alpha$/H$\beta$ disparity
takes place in early spectra of SN~2008in both on day 11 and day 18.
It is shown that the problem can be solved in terms of the clumpy
density structure in the outer SN layers with a mass of $\sim 0.03~M_{\odot}$.
The clumpy structure of the outer layers of 
supernova is suggested to be an outcome of the density perturbation in 
the RSG external convective zone combined with the density inversion.
The shock propagation in this medium presumably generates inhomogeneities 
in SN ejecta with a high density contrast. 

Tests are proposed for the verification of the clumpy
density distribution of outer layers of SN~IIP. 
The first is based on the detection of the flux fluctuations in 
the early H$\alpha$ line profile at the level of $\leq 1$\%. 
The second, indirect, test suggests the analysis of the properties 
of the initial radiation outburst related to the shock breakout.

\acknowledgements
We thank R. Roy for kindly sending us spectra of SN~2008in.

{\bf References}
\bigskip

\noindent G. Aldering, R. M. Humphreys, and M. Richmond,
Astron. J. {\bf107}, 662 (1994)\\
P. V. Baklanov, S. I. Blinnikov, and N. N. Pavlyuk,
Astron. Lett. {\bf31}, 429 (2005)\\
E. Baron, P. H. Hauschildt, D. Branch, et al.,
Astrophys. J. {\bf441}, 170 (1995)\\
A. Chiavassa, B. Freytag, T. Masseron, and B. Plez,
 Astron. Astrophys. {\bf535A}, 22 (2011)\\
N. N. Chugai, Astrophys. J. {\bf428}, L17 (1994)\\
N. N. Chugai, Mon. Not. Roy. astr. Soc. {\bf326}, 1448 (2001)\\
L. Dessart, S. Blondin, P. Brown, et al., Astrophys. J. {\bf675}, 644 (2008)\\
R. G. Eastman, S. E. Woosley, T. A. Weaver, and Ph. A. Pinto, Astrophys. J. \\
\indent {\bf430}, 300 (1994)\\
Yu. A. Fadeyev, Astron. Lett. {\bf38}, 260 (2012)\\
S. W. Falk and W. D. Arnett, Astrophys. J. Suppl. {\bf33}, 515 (1977)\\  
E. K. Grassberg, V. S. Imshennik, and D. K. Nadyozhin, Astrophys. Space Sci. {\bf10}, \\ 
\indent 28 (1971)\\
A. Heger, A., C. L. Fryer, S. E. Woosley, N. Langer, N., and D. H. Hartmann,\\
\indent Astrophys. J. {\bf591}, 288 (2003) \\
L. Hesselink and B. Sturtevant, J. Fluid Mech. {\bf196}, 513 (1988)\\
M. G. Kendall and P. A. P. Moran, {\it Geometrical Probability} (New York: Hafner, 1963) 
 p. 90\\
F. Kupka, N. Piskunov, T. A. Ryabchikova, et al., Astron. Astrophys. Suppl.\\
\indent  {\bf138}, 119 (1999)\\
L. D. Landau and E. M. Lifshitz, {\it Fluid mechanics} (Oxford: Pergamon Press 1987)
 p. 21\\
D. C. Leonard, A. V. Filippenko, E. L. Gates, et al., Publ. Astr. Soc. Pacific \\
\indent  {\bf114}, 35 (2002)\\
A. Maeder, Proceedings of the International Colloquium, Amsterdam, 26 February - 1 March \\
\indent 1991 (Ed. C. de Jager, H. Nieuwenhuijzen,  Amsterdam: North-Holland, 1992), p.138\\
R. M. Quimby, J. C. Wheeler, P. Hoeflich. et al., Astrophys. J. {\bf666}, 1093 (2007) \\
R. Roy, B. Kumar, S. Benetti, et al., Astrophys. J. {\bf736}, 76 (2011)\\
L. Tomasella, E. Cappellaro, M. Fraser, et al., Mon. Not. Roy. astr. Soc. {\bf 434}, \\
\indent 1636 (2013)\\
V. Utrobin, Astron. Astrophys. {\bf461}, 233 (2007)\\
V. Utrobin and N. Chugai (UC13),  Astron. Astrophys. {\bf461}, 233 (2013)\\

\clearpage

\begin{table}
  \caption{Clumpy models on days 11 and 18}
  \begin{tabular}{cccccccccc}
\hline

Model & $v_p(\rm{H}\alpha)$ & $v_p(\rm{H}\beta)$ & $v_f$ & $\tau_c$  
 & $\tau_{ic}$ &  $p$ & $q$ & $f_0$ & $k$ \\ 
   & \multicolumn{3}{c}{km s$^{-1}$} &    &     &     &     &       &  \\      
\hline

 D11 & 6100 & 6100 & 6100 & 100  &  2   &  -5.5   &  0   & 0.55  &  -2  \\
 D18 & 4100 & 5300 & 6100 & 100  &  1.5   &  -5.5   &  0   & 0.55  & -2  \\
\hline

\end{tabular}
\label{t-mod} 
\end{table} 
\clearpage

 \begin{figure}
\plotone{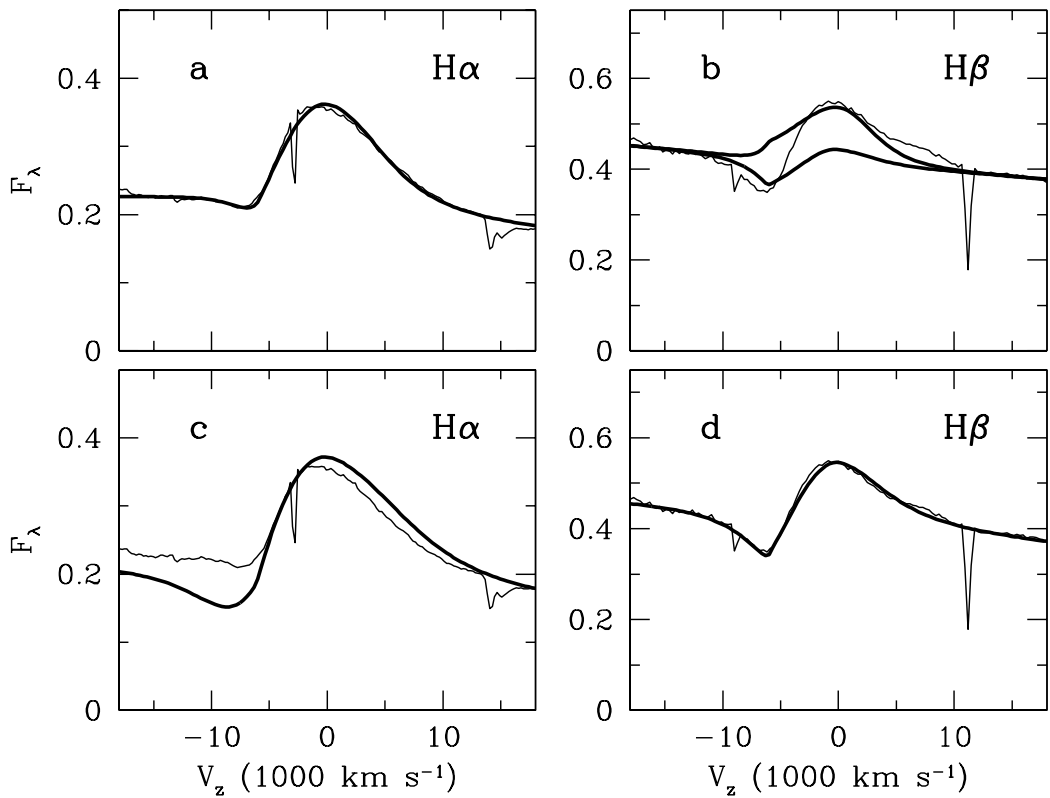}
\caption{
H$\alpha$ and H$\beta$ lines in SN~2008in spectrum on day 11 (thin line)
compared to models of the spherical smooth supernova atmosphere.
Panel {\bf a} is the case with parameters adjusted to fit H$\alpha$;
panel {\bf b} is the model H$\beta$ for the same population of the hydrogen 
second level as for H$\alpha$ without net emission (lower line) and 
with the net emission (upper line);
panel {\bf c} is the model H$\alpha$ for the second level population 
adjusted for the H$\beta$ (panel {\bf d}). The plots demonstrate 
the obvious disparity between H$\alpha$ and H$\beta$ in the model of the 
spherical smooth ejecta.}
\end{figure}
\clearpage 

\begin{figure}
\plotone{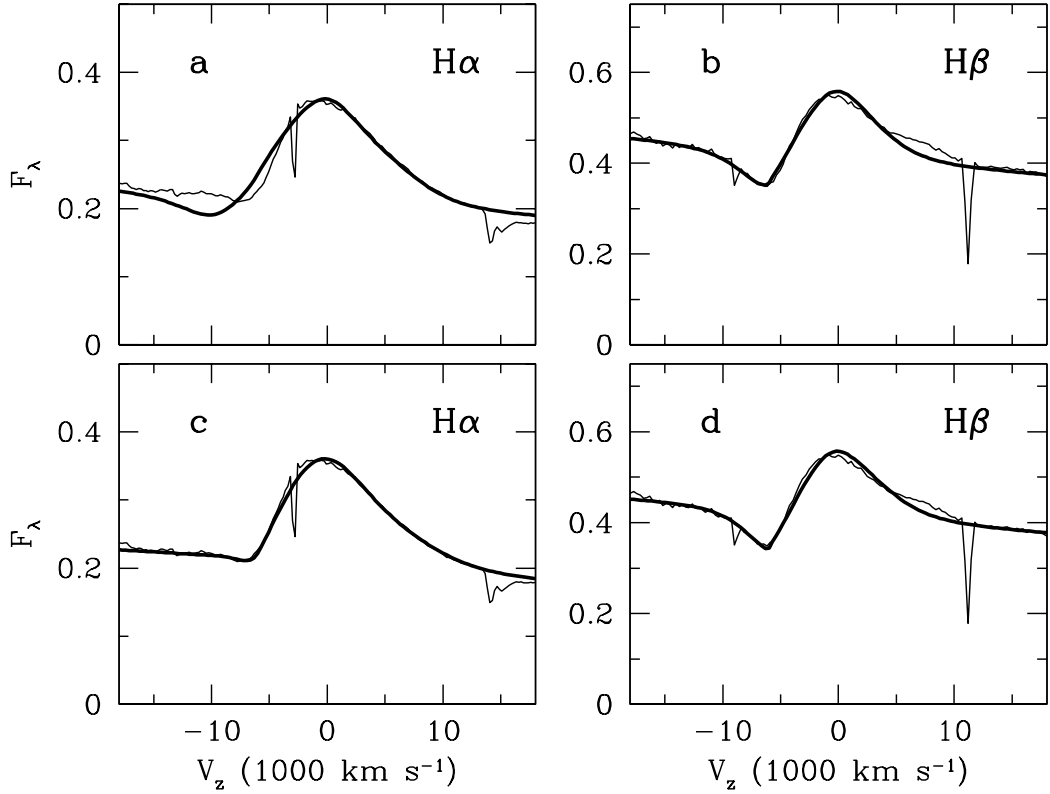}
\caption{
The same as in Fig. 1 but with the models of clumpy envelope.
Panels {\bf a} and {\bf b} show the model with a constant 
filling factor in the atmosphere. This model fails to describe 
simultaneously the H$\alpha$ and H$\beta$ lines. 
Panels {\bf c} and {\bf d} present the model with the filing factor 
decreasing outward. The model successfully reproduces both lines.}
\end{figure}
\clearpage 

\begin{figure}
\plotone{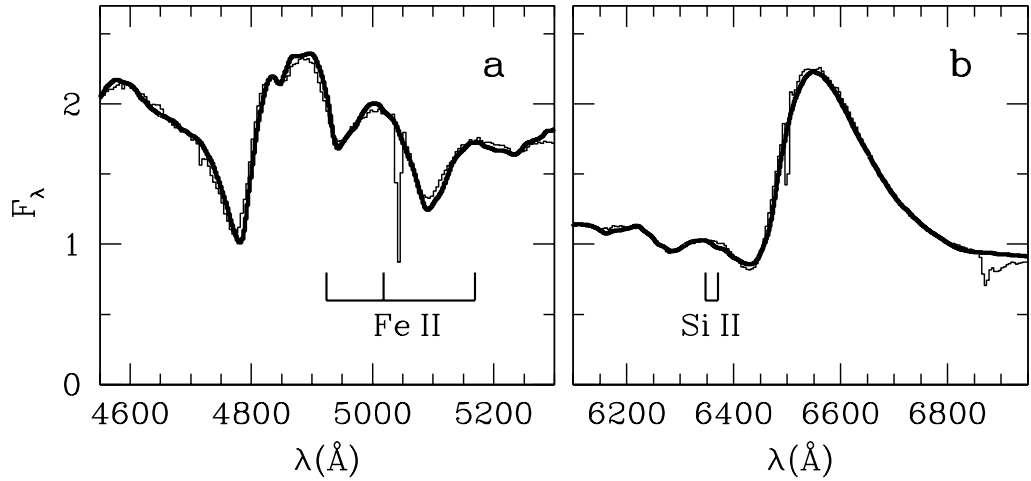}
\caption{
H$\beta$ (panel {\bf a}) and H$\alpha$ (panel {\bf b}) bands in the SN~2008in 
spectrum (thin line) on day 18 compared to the model of smooth
atmosphere in which the metal lines are taken into account.
Shown also are the positions of strongest Fe\,II lines 
(multiplet 42) and Si\,II 
doublet which contribute to the H$\beta$ and H$\alpha$ bands respectively.
Note, in the H$\beta$ model population of the second level is three times 
larger than that for the H$\alpha$ model. This demonstrates that the 
disparity between the H$\alpha$ and H$\beta$ exists on day 18 as well.}
\end{figure}
\clearpage

\begin{figure}
\plotone{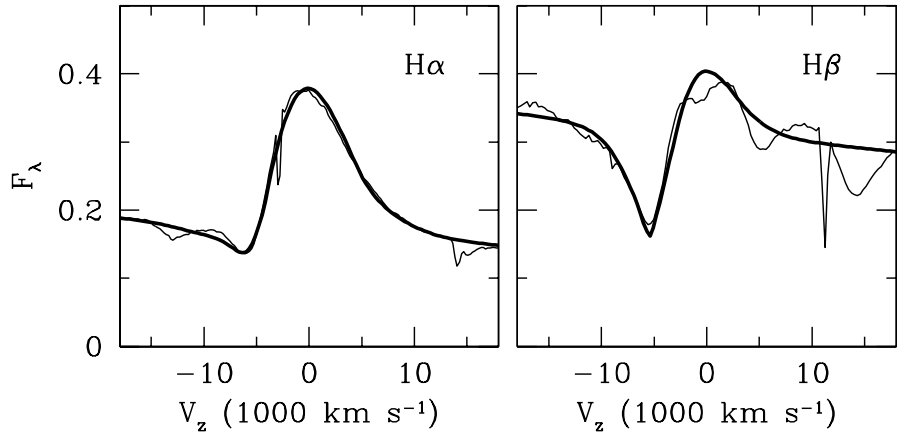}
\caption{
H$\alpha$ and H$\beta$ in the SN~2008in spectrum on day 18 (thin line) 
compared to the clumpy model that was applied on day 11. The 
successful description of both lines indicates that 
between day 11 and 18 the clumpy structure evolves homologously.}
\end{figure}
\clearpage

\begin{figure}
\plotone{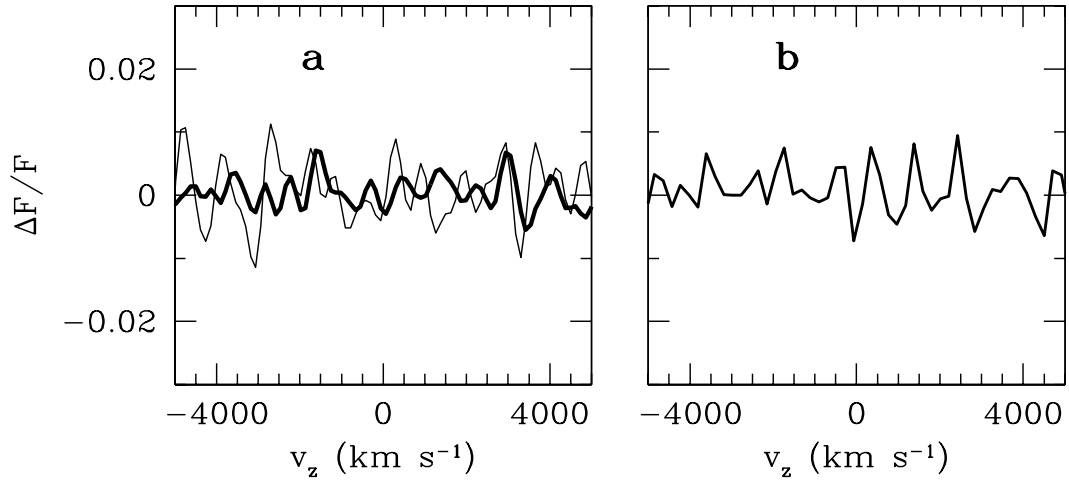}
\caption{
Residual spectrum of H$\alpha$ in the model of cloudy envelope 
(panel {\bf a}) for cloud radius of 100 km s$^{-1}$ (thick line) and 
250 km s$^{-1}$ (thin line) on day 11. Panel {\bf b} is the residual 
spectrum in the same band for the observed H$\alpha$ line on day 11.
The residual spectrum is the spectrum divided on the heavily 
smoothed spectrum with the subsequent subtraction of unity.}
\end{figure}

\end{document}